\documentclass[preprint,aps,prl,showpacs,superscriptaddress]{revtex4}

\newenvironment{eq}{\begin{equation}}{\end{equation}}
\newenvironment{ea}{\begin{eqnarray}}{\end{eqnarray}}
\usepackage{graphics}
\usepackage{bm}
\begin{document}
\title{Phonon Dispersion in Chiral Single Wall Carbon Nanotubes}
\author{Weihua \surname{MU}}
\email[Email address: ] {muwh@itp.ac.cn} \affiliation{Institute of
Theoretical Physics,
 The Chinese Academy of Sciences,
 P.O.Box 2735 Beijing 100080, China}
\author{Anthony Nickolas \surname{Vamivakas}}
\affiliation{Department of Electrical and Computer 
Engineering, Boston University, 8 St. Mary's Stret Boston MA 02215, USA}
\author{Zhong-can \surname{Ou-Yang}}
\affiliation{Institute of Theoretical Physics,
 The Chinese Academy of Sciences,
 P.O.Box 2735 Beijing 100080, China}
\affiliation{Center for Advanced Study,
 Tsinghua University, Beijing 100084, China}
\begin{abstract}
%%%%%%%%%%%%%%%%%%%%%%%%%%%%%%%%%%%%%
The phonon dispersion of chiral single wall carbon nanotubes has
been obtained from $6\times 6$ dyanimic matrix. The present manuscript is the extension of G. D.
Mahan and Gun Sang Jeon's work on armchair and zigzag nanotubes.
[see P. R. B., {\bf 70}, 075405~(2004)]. We use spring
and mass model with the proper phonon potential suggested by
\cite{M2}. We can calculate the dispersion of single wall carbon
nanotubes near $\Gamma$ point with arbitrary chirality. The
results are compatible with Mahan et. al's results for armchair
and zigzag SWNTs.
\end{abstract}

 \maketitle

Carbon nanotubes have attracted much interest on their electronic
and thermal transport properties \cite{De}. Single wall carbon nanotubes
(SWNTs) can be thought of as a sheet of graphite rolled into a cylinder
\cite{De}. In many experiments, phonons play an important role \cite{kong,Nick},
especially, many of properties of SWNTs are determintd by the low-energy
phonon excitations. Therefore there is a pressing need for understanding
the phonon dispersion in carbon nanotubes, There are several works
on this topic \cite{Ye,popov,Dobard,Daniel,Mounet, Dubay,1,2,3,4,5,6,9}.
% Recently, L-H. Ye, \emph{et,
%al.\cite{Ye}} use \emph{ab initio} supercell approach to obtain th
%phonon-dispersion curves for a chiral SWNTs, and find there is no
%soft phonon mode presented in low-frequency region. 
G. D. Mahan and Gun Sang Jeon proposed a tractable
method in terms of spring and mass model to calculate the phonon dispersion
of armchair and zigzag SWNTs \cite{M2}.

In the present paper, we investigate theoretically the phonon dispersion
of SWNTs with arbitrary chiral index (n,m). We follow the algorithm
proposed in \cite{M2} and we extend it for our calculations.

In SWNT, there are two carbon atoms in a unit cell. We label A atoms
at the lattice site $\mathbf{R}_{j}$ and their three nearest neighbors
B atoms at site $\mathbf{R}_{j}+\boldsymbol \delta $. Unit vector
$\hat{\boldsymbol \delta }$ reflects the direction of $\boldsymbol \delta $.

\begin{figure}[http]
\scalebox{1.0}{\includegraphics{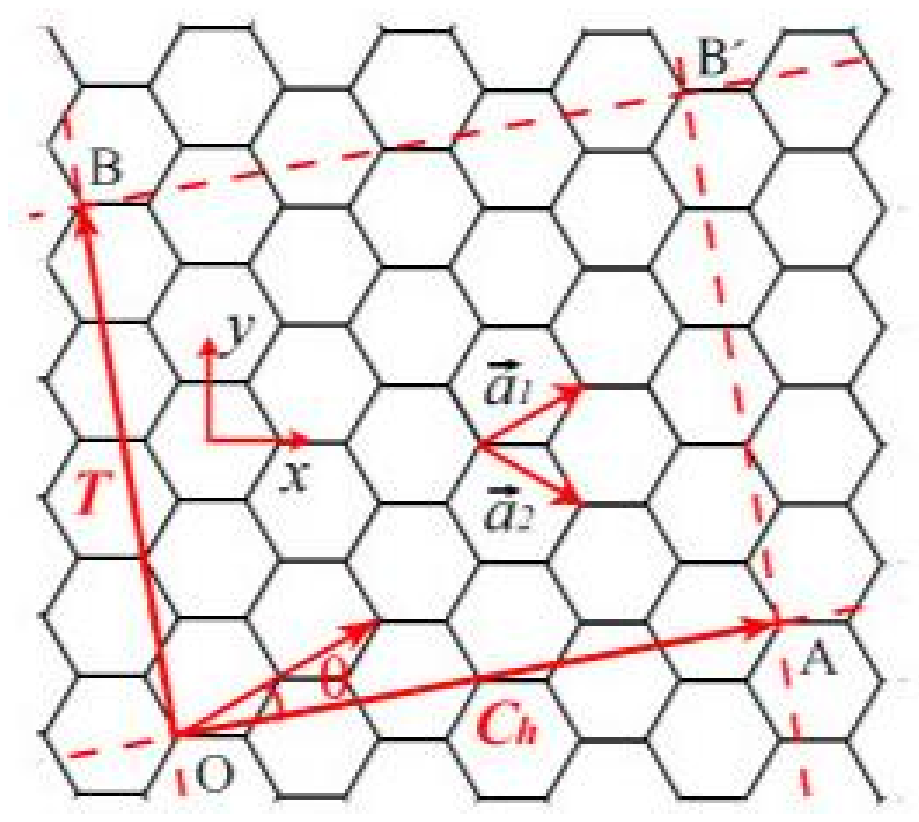}}
\caption{\label{Fig1} The 2D sketch map of single wall carbon
nanotubes. This image was made with VMD and is owned by the
Theoretical and Computational Biophysics Group, an NIH Resource
for Macromolecular Modeling and Bioinformatics, at the Beckman
Institute, University of Illinois at Urbana-Champaign.}
\end{figure}

Fig.\ref{Fig1} shows the commonly used coordinate system in nanotubes
with base vectors \begin{eqnarray*}
\vec{a}_{1} & = & \sqrt{3}a(\sqrt{3}/2,\; 1/2),\\
\vec{a}_{2} & = & \sqrt{3}a(\sqrt{3}/2,-1/2),,
\end{eqnarray*}

The chirality of a certain carbon nanotube is characterized by a pair
of integer (n,m). When a sheet of graphite is rolled into a cylinder,
the points (0,0) and (n,m) are identical, and the diameter of tube
$R$ is the distance between above two points on the 2D plane divided
by $2\pi $.

Consider a type A Atom located at lattice site (t,l) on 2D graphite
sheet. For an (n,m) SWNT with radius $R$, the atom position in 3D
Descartes coordinate system is, ($c=3a/2,\theta _{z}=\sqrt{3}a/R$)
\begin{eqnarray}
R_{A} & = & [R\cos (\theta _{tl}),R\sin (\theta _{tl}),\bar{l}c],\label{Coor}\\
R_{B}^{(i)} & = & [R\cos (\theta _{tl}+u_{i}\theta _{z}),R\sin (\theta _{tl}+u_{i}\theta _{z}),\bar{l}c+av_{i}],
\end{eqnarray}
 where \begin{equation}
u_{i}=\left\{ \begin{array}{ccc}
 \frac{n+m}{2\sqrt{n^{2}+nm+m^{2}}},\hspace {1cm}i=1, &  & \\
 \newline \frac{-n}{2\sqrt{n^{2}+nm+m^{2}}},\hspace {1cm}i=2, &  & \\
 \newline \frac{-m}{2\sqrt{n^{2}+nm+m^{2}}},\hspace {1cm}i=3. &  & \end{array}
\right.\end{equation}
 and \begin{equation}
v_{i}=\left\{ \begin{array}{cc}
 \frac{n-m}{2\sqrt{n^{2}+nm+m^{2}}},\hspace {1cm}i=1, & \\
 \newline \frac{n+2m}{2\sqrt{n^{2}+nm+m^{2}}},\hspace {1cm}i=2, & \\
 \newline -\frac{2n+m}{2\sqrt{n^{2}+nm+m^{2}}},\hspace {1cm}i=3. & \end{array}
\right.\end{equation}
 Where $\bar{l}=\frac{\sqrt{3}}{2}\cdot \frac{ln-tm}{\sqrt{n^{2}+nm+m^{2}}}$
denotes the position along the $z$ axis.

It is easy to verify that the carbon atoms' coordinates of armchair
tube $m=n$ and zigzag tube $m=0$ shown in ref. \cite{M1,M2} are
just special examples of our general expression.

According to the atom coordinates given above, the three unit vectors
$\hat{\delta }_{i}$ are \begin{equation}
\hat{\delta }^{(i)}=[-\sqrt{3}u_{i}\sin (\theta _{tl}+\frac{u_{i}}{2}\theta _{z}),\sqrt{3}u_{i}\cos (\theta _{tl}+\frac{u_{i}}{2}\theta _{z}),v_{i}],\hspace {1cm}i=1,2,3.\label{delta}\end{equation}

We know that for carbon nanotubes, the basis of lattice vectors in
real space, in cylindical coordinate system, \cite{1,tu}

\begin{eqnarray*}
\vec{\alpha }_{1} & = & (\frac{2\pi }{N},\tau )\equiv (\psi ,\tau ),\\
\vec{\alpha }_{2} & = & (0,T).
\end{eqnarray*}

The corresponding reciprocal lattice vectors are

\begin{eqnarray*}
\vec{\beta }_{1} & = & (N,0),\\
\vec{\beta }_{2} & = & (-\frac{\tau N}{T},\frac{2\pi }{T}).
\end{eqnarray*}

Here, \begin{eqnarray*}
N & = & \frac{2(n^{2}+nm+m^{2})}{d_{R}},\\
T & = & \frac{3a\sqrt{n^{2}+nm+m^{2}}}{d_{R}},\\
d_{R} & \equiv  & gcd(n,m).
\end{eqnarray*}

It is easy to verify that, 

\[
\vec{\alpha }_{i}\cdot \vec{\beta }_{j}=2\pi \delta _{ij}.\]

In SWNTs, the electron states have two quantum numbers $(q,\alpha )$,
where continuous index $q$ is the wave vector along the tube axis.
Discrete quantum number $\alpha $ reflects the angular dependence
around the tube. In first Brillouin zone, then, \[
\alpha =0,1,...,N-1,\quad -\frac{\pi }{T}\leq q\leq \frac{\pi }{T}.\]

In cylinder coordinate system $[\rho ,\theta ,z]$, we consider the
displacement of the atoms A and B in three directors ${Q_{s\rho },Q_{s\theta },Q_{sz}},(s=A,B)$
are, \begin{eqnarray}
Q_{A,tl} & = & Q_{A,tl,\rho }[\cos (\theta _{tl}),\sin (\theta _{tl}),0]\nonumber \\
 & + & Q_{A,tl,\theta }[-\sin (\theta _{tl}),\cos (\theta _{tl}),0]+Q_{A,tl,z}[0,0,1],\\
Q_{B,tl}^{(i)} & = & Q_{B,tl,\rho }[\cos (\theta _{tl}+u_{i}\theta _{z}),\sin (\theta _{tl}+u_{i}\theta _{z}),0]\nonumber \\
 & + & Q_{B,tl,\theta }[-\sin (\theta _{tl}+u_{i}\theta _{z},\cos (\theta _{tl}+u_{i}\theta _{z}),0]\nonumber \\
 & + & Q_{B,tl,z}[0,0,1],\hspace {1cm}i=1,2,3.
\end{eqnarray}

The wave vectors occur in the collective coordinates. Wave vector
$q$ is used for the tube axis, the other quantum number $\alpha $
denotes the position around the circumference of the tube. Similar
to the discussion for the single electron dispersion, we have $\alpha =0,1...,N-1$
with $N=\frac{2(n^{2}+nm+m^{2})}{d_{R}}$. Thus, \begin{eqnarray*}
Q_{A,tl,r} & = & \frac{1}{\sqrt{N_{a}}}\sum _{q\alpha }Q_{A,q\alpha ,r}\exp [i(qc\bar{l})+\alpha \theta _{tl}],\\
Q_{B,tl,r}^{(i)} & = & \frac{1}{\sqrt{N_{a}}}\sum _{q\alpha }Q_{B,q\alpha ,r}\exp [i(qc\bar{l}+iqav_{i})+\alpha (\theta _{tl}+u_{i}\theta _{z})],
\end{eqnarray*}
 where $r=(\rho ,\theta ,z)$ and $N_{a}$ is the total number of
carbon atoms in SWNTs.

The phonon potential energy $V_{1}$ is \cite{M2}\begin{eqnarray}
V_{1} & = & \frac{K_{1}}{2}\sum _{tl}\sum _{i=1}^{3}|\Xi ^{(i)}|^{2},\\
\Xi ^{(i)} & = & \hat{\delta }_{i}\cdot (\vec{Q}_{B}^{(i)}-\vec{Q}_{A})\nonumber \\
 & = & e^{i(qcl+\beta \theta _{tl})}\tilde{\Xi }^{(i)},i=1,2,3\\
\tilde{\Xi }^{(i)} & = & e^{iqav_{i}+i\alpha u_{i}\theta _{z}}(\sqrt{3}u_{i}\sin (\frac{u_{i}\theta _{z}}{2})Q_{B,\rho }+\sqrt{3}u_{i}\cos (\frac{u_{i}\theta _{z}}{2})Q_{B\theta }+v_{i}Q_{Bz})\nonumber \\
 & + & \sqrt{3}u_{i}\sin (\frac{u_{i}\theta _{z}}{2})Q_{A\rho }-\sqrt{3}u_{i}\cos (\frac{u_{i}\theta _{z}}{2})Q_{A\theta }-v_{i}Q_{Az},i=1,2,3.
\end{eqnarray}

The displacement $Q$ is a complex quantity, and the real and imaginary
displacements are independent variables. According to ref. \cite{M2},
we only need the force on $Q$ as $\delta V/\delta Q^{*}$ to obtain
corresponding determinant equation, \begin{eq}
0=det\left|
\begin{array}{cccccc}
M_{11}-\Omega_{1}^{\;2} & M_{12} & M_{13} & M_{14} & M_{15} & M_{16}\\
(M_{12})^{\;*} & M_{22}-\Omega_{1}^{\;2} & M_{23} & M_{24} & M_{25} & M_{26}\\
(M_{13})^{\;*} & (M_{23})^{\;*} & M_{33}-\Omega_{1}^{\;2} & M_{34} & M_{35} & M_{36}\\
(M_{14})^{\;*} & (M_{24})^{\;*} & (M_{34})^{\;*}& M_{44}-\Omega_{1}^{\;2}  & M_{45} & M_{46}\\
(M_{15})^{\;*} & (M_{25})^{\;*} & (M_{35})^{\;*} & (M_{45})^{\;*} & M_{55}-\Omega_{1}^{\;2} & M_{56}\\
(M_{16})^{\;*} & (M_{26})^{\;*} & (M_{36})^{\;*} & (M_{46})^{\;*} &
(M_{56})^{\;*} & M_{66}-\Omega_{1}^{\;2}
\end{array}
\right|,
\end{eq} where \begin{ea}
M_{11}&=&\sum_{i=1}^{3} 3u_{i}^{2} s_{iz}^{2},\;\;\;\;\;\;
M_{12}=\sum_{i=1}^{3} -3u_{i}^{2}
s_{iz}c_{iz},\;\;\;\;\;\;\;\; M_{13}=\sum_{i=1}^{3} -\sqrt{3}u_{i} v_{i} s_{iz}, \nonumber\\
M_{14}&=&\sum_{i=1}^{3} 3u_{i}^{2} s_{iz}^{2}
e^{i\beta_{i}},\;M_{15}=\sum_{i=1}^{3} 3u_{i}^{2} s_{iz}c_{iz}
e^{i\beta_{i}},\; M_{16}=\sum_{i=1}^{3} \sqrt{3}u_{i} v_{i} s_{iz}
e^{i\beta_{i}},\nonumber\\
M_{22}&=&\sum_{i=1}^{3} 3u_{i}^{2} c_{iz}^{2}-\Omega^{2}
,\;\;\;\;\;M_{23}=\sum_{i=1}^{3} \sqrt{3}u_{i} v_{i} c_{iz},
\;\;\;\;M_{24}=\sum_{i=1}^{3} - 3u_{i}^{2} s_{iz}c_{iz} e^{i\beta_{i}},\nonumber\\
M_{25}&=&\sum_{i=1}^{3} - 3u_{i}^{2} c_{iz}^{2}
e^{i\beta_{i}},\;M_{26}=\sum_{i=1}^{3} - \sqrt{3}u_{i} v_{i} c_{iz}
e^{i\beta_{i}},\;M_{33}=\sum_{i=1}^{3} v_{i}^{2}-\Omega^{2},\nonumber\\
M_{34}&=&\sum_{i=1}^{3} -\sqrt{3}u_{i} v_{i}
s_{iz}e^{i\beta_{i}},\;M_{35}=\sum_{i=1}^{3} -\ \sqrt{3}u_{i} v_{i} c_{iz}
e^{i\beta_{i}},\;M_{36}=\sum_{i=1}^{3} - v_{i}^{2} e^{i\beta_{i}},
\nonumber\\
M_{44}&=&\sum_{i=1}^{3} \ 3u_{i}^{2} s_{iz}^{2}
-\Omega^{2},\;\;\;\;M_{45}=\sum_{i=1}^{3} 3u_{i}^{2}
s_{iz}c_{iz},\;\;\;\;M_{46}=\sum_{i=1}^{3} \sqrt{3}u_{i} v_{i} s_{iz},
\nonumber\\
M_{55}&=&\sum_{i=1}^{3} 3u_{i}^{2} c_{iz}^{2}
-\Omega^{2},\;\;\;\;\;M_{56}=\sum_{i=1}^{3} \sqrt{3}u_{i} v_{i}
c_{iz},\;\;\;\; M_{66}=\sum_{i=1}^{3} -v_{i}^{2}, \nonumber
\end{ea} where $s_{iz}\equiv \sin (\frac{u_{i}\theta _{z}}{2})$ , $c_{iz}\equiv \cos (\frac{u_{i}\theta _{z}}{2})$
and $\beta _{i}\equiv qav_{i}+\alpha u_{i}\theta _{z}$. For convenience,
a dimensionless quantity $\Omega _{1}^{\; 2}=m\omega ^{2}/K_{1}$
is defined. $K_{1}$ is the spring constant.

If the SWNT is zigzag or armchair type, our results are reduced to
the results shown in ref. \cite{M2}. The above determinant equation
has analytical solutions when $q=0$. Since $\alpha $ has $N$ possible
values, the phonon modes of chiral SWNTs' number is much more than
achiral SWNTs'. The calculated results are unphysical, for there are
three zero-frequency modes and have no acoustical modes or flexure
modes. Following ref. \cite{M2}'s discussion, we have to consider
the second neighbor interaction.

We thus use springs between second neighbors. For A atoms with coordinate
$R_{A}$, its six second neighbors are with coordinates \begin{eqnarray}
R_{A}^{(1,2,3)} & = & [R\cos (\theta _{tl}+v_{i}\theta _{z}),R\sin (\theta _{tl}+v_{i}\theta _{z}),\bar{l}c+3au_{i}],i=1,2,3\label{Coor1}\\
R_{A}^{(4,5,6)} & = & [R\cos (\theta _{tl}-v_{i}\theta _{z}),R\sin (\theta _{tl}-v_{i}\theta _{z}),\bar{l}c-3au_{i}],i=1,2,3
\end{eqnarray}
 and the unit vectors to them are \begin{ea}
\label{delta1}
\widetilde{\hat{\delta}}^{(i)}&=&[-v_{i}\sin(\theta_{tl}+\frac{v_{i}}{2}\theta_{z}),
u_{i}\cos(\theta_{tl}+\frac{u_{i}}{2}\theta_{z}),
\sqrt{3}u_{i}],\hspace{1cm}i=1,2,3\nonumber\\
\widetilde{\hat{\delta}}^{(i+3)}&=&-\widetilde{\hat{\delta}}^{(i)},\hspace{7.9cm}
i=1,2,3
\end{ea} We get \begin{ea}
\widetilde{\tilde{\Xi}}^{(i)} &=&e^{i3qau_{i}+i\alpha v_{i}\theta
_{z}}[v_{i} \sin(\frac{v_{i}
\theta_{z}}{2})Q_{A,\rho}+v_{i}\cos(\frac{v_{i} \theta_{z}}
{2})Q_{A\theta }+\sqrt{3}u_{i}Q_{Az}]  \nonumber \\
&+&v_{i}\sin(\frac{v_{i} \theta_{z}}{2})Q_{A\rho }-v_{i}
\cos(\frac{v_{i} \theta_{z}}{2})Q_{A\theta
}-\sqrt{3}u_{i}Q_{Az},~~~i=1,2,3.\nonumber\\
\widetilde{\tilde{\Xi}}^{(i+3)} &=&e^{-i3qau_{i}-i\alpha v_{i}\theta
_{z}} [v_{i}\sin(\frac{v_{i} \theta_{z}}{2})Q_{A,\rho}-
v_{i}\cos(\frac{v_{i} \theta_{z}}{2})Q_{A\theta }-\sqrt{3}u_{i}Q_{Az}]
 \nonumber \\
&+&v_{i}\sin(\frac{v_{i} \theta_{z}}{2})Q_{A\rho }+
v_{i}\cos(\frac{v_{i} \theta_{z}}{2})Q_{A\theta
}+\sqrt{3}u_{i}Q_{Az},~~~i=1,2,3.
\end{ea}

We reduce above equations to \begin{ea}
\widetilde{\tilde{\Xi}}^{(i)}
&=&2e^{i\tilde{\beta}_{i}/2}\{v_{i}\sin(\frac{v_{i}
\theta_{z}}{2})Q_{A,\rho}\cos{\tilde{\beta}_{i}/2}+i[v_{i}
\cos(\frac{v_{i}
\theta_{z}} {2})Q_{A\theta }\nonumber\\
&+&\sqrt{3}u_{i}Q_{Az}]\sin(\tilde{\beta}_{i}/2)\}
\hspace{4.5cm}i=1,2,3,
\nonumber \\
\widetilde{\tilde{\Xi}}^{(i+3)} &=&
(\widetilde{\tilde{\Xi}}^{(i)})^{*} \hspace{6.6cm}i=1,2,3.
\end{ea} where $\tilde{\beta }_{i}=3qau_{i}+\alpha \theta _{z}v_{i}$ and
the $z^{*}$ denotes the complex conjugate of $z$. Obviously, $|\widetilde{\tilde{\Xi }}^{(i+3)}|=|\widetilde{\tilde{\Xi }}^{(i)}|$.
These interactions give a dynamical matrix for the oscillation of
the A sublattice \begin{eq}
0=det\left|
\begin{array}{ccc}
8\sum_{i=1}^{3} v_{i}^{2} \widetilde{s_{iz}}^{2}C_{i}^{2}-\Omega_{2}^{\;2} &
8\sum_{i=1}^{3} iv_{i}^{2} \widetilde{s_{iz}}\widetilde{c_{iz}}S_{i}C_{i}
& 8\sum_{i=1}^{3} i\sqrt{3}u_{i} v_{i}
s_{iz}S_{i}C_{i} \\
-8\sum_{i=1}^{3} iv_{i}^{2}
\widetilde{s_{iz}}\widetilde{c_{iz}}S_{i}C_{i}& 8\sum_{i=1}^{3} v_{i}^{2}
\widetilde{c_{iz}}^{2}S_{i}^{2}-\Omega_{2}^{\;2}&8\sum_{i=1}^{3} \sqrt{3}u_{i}
v_{i}
c_{iz}S_{i}^{2}\\
-8\sum_{i=1}^{3} i\sqrt{3}u_{i} v_{i} \widetilde{s_{iz}}S_{i}C_{i}
&8\sum_{i=1}^{3} \sqrt{3}u_{i} v_{i} \widetilde{c_{iz}}S_{i}^{2}
&8\sum_{i=1}^{3} 3u_{i}^{2}S_{i}^{2}-{\Omega_{2}}^{\;2}
\end{array}
\right|
\end{eq} where, $\widetilde{s_{iz}}\equiv \sin (\frac{v_{i}\theta _{z}}{2}),\widetilde{c_{iz}}\equiv \cos (\frac{v_{i}\theta _{z}}{2}),S_{i}\equiv \sin (\frac{\tilde{\beta }_{i}}{2})$
and $C_{i}\equiv \cos (\frac{\tilde{\beta }_{i}}{2})$. The dimensionless
quantity $\Omega _{2}^{\; 2}\equiv M\omega ^{2}/K_{2}$. $K_{2}$
is the spring constant.

An identical set of equations applies to the second neighbor interactions
among the B atoms. From the above determinant equation, we find there
are two acoustical modes at $\alpha =0$, i.e., longitudinal acoustical
phonon and the twist mode. From corresponding velocities of the sound
$v_{L}$ and $v_{TW}$, we may fit the spring constants \cite{De}.
In addition, when $\alpha =1$, there occurs one flexure mode. It
implies that we should consider radial bond bending effect to get
ZO phonon modes, which is similar to the out-of-plane mode of graphene.

For a sheet of graphene, the vibrations perpendicular to the plane
are well described by bond-bending forces.For SWNTs, in order to investigate
the radial bond bending, ref. \cite{M2} gives the relative phonon
potential energy \begin{ea}
V_{3}&=&\frac{K_{3}}{2}\sum_{j} |\Delta_{j}|^{2},\\
|\Delta_{j}|&=&\sum_{i=1}^{3} \hat{n}_{i} \cdot
(\vec{Q}_{j}^{(i)}-\vec{Q}_{j}),
\end{ea} where $\vec{Q}_{j}^{(i)}$ are the first neighbors of $\vec{Q}_{j}$.
If $\vec{Q}_{j}$ denotes an A atom, the normal vectors are \begin{eq}
\vec{n}_{i}=[\cos(\theta_{tl}+\frac{u_{i}}{2}\theta_{z}),\sin(\theta_{tl}+\frac{u_{i}}{2}\theta_{z}),0],
\hspace{1cm} i=1,2,3.
\end{eq} In polar coordinates, using the Fourier transformations, these two
interactions are \begin{ea}
\Delta_{A}&=&e^{iqav_{i}+i\alpha u_{i}\theta_{z}}[Q_{B\rho}\cos(\frac{u_{i} \theta_{z}}{2})-Q_{B\theta}\sin(\frac{u_{i} \theta_{z}}{2})]\nonumber\\
&-&[Q_{A\rho}\cos(\frac{u_{i} \theta_{z}}{2})+Q_{A\theta}\sin(\frac{u_{i} \theta_{z}}{2})]\\\Delta_{B}&=&e^{-iqav_{i}-i\alpha u_{i}\theta_{z}}[Q_{A\rho}\cos(\frac{u_{i} \theta_{z}}{2})+Q_{A\theta}\sin(\frac{u_{i} \theta_{z}}{2})]\nonumber\\
&-&[Q_{B\rho}\cos(\frac{u_{i} \theta_{z}}{2})-Q_{B\theta}\sin(\frac{u_{i}
\theta_{z}}{2})]
\end{ea} The interaction leads to a dynamical matrix \begin{eq}
0=det\left|
\begin{array}{cccc}
c^{2}+|\widetilde{c}|^{2}-\Omega_{3}^{\;2} &
cs+\widetilde{c}\widetilde{s}^{\;\;*} & -2 c\widetilde{c}& s1
\widetilde{c}+c
\widetilde{s}\\
c s+\widetilde{s}\widetilde{c}^{\;\;*}&
s^{2}+|\widetilde{s}|^{2}-\Omega_{3}^{\;2}& -s \widetilde{c}-c
\widetilde{s} &
2s \widetilde{s}\\
-2 c\widetilde{c}^{\;\;*}& -s\widetilde{c}^{\;\;*}-c
\widetilde{s}^{\;\;*}& c^{2}+|\widetilde{c}|^{2}-\Omega_{3}^{\;2}&-c
s-\widetilde{s}\widetilde{c}^{\;\;*}\\
s \widetilde{c}^{\;\;*}+c \widetilde{s}^{\;\;*}& 2 s
\widetilde{s}^{\;\;*}& -c s-\widetilde{c} \widetilde{s}^{\;\;*}&
s^{2}+|\widetilde{s}|^{2}-\Omega_{3}^{\;2}
\end{array}
\right|
\end{eq} where $c\equiv \sum _{i=1}^{3}\cos (\frac{u_{i}\theta _{z}}{2})$
, $\widetilde{c}\equiv \sum _{i=1}^{3}\cos (\frac{u_{i}\theta _{z}}{2})e^{i\beta _{i}}$,
$s\equiv \sum _{i=1}^{3}\sin (\frac{u_{i}\theta _{z}}{2})$ and $\widetilde{s}\equiv \sum _{i=1}^{3}\sin (\frac{u_{i}\theta _{z}}{2})e^{i\beta _{i}}$.
The dimensionless quantity $\Omega _{3}^{\; 2}=M\omega ^{2}/2K_{3}$.
$K_{3}$ is the spring constant.

When $\alpha =0$, the solution of above equation does have ZO modes.
At $q=0$, $\omega _{ZO}$ is also important to fit spring constants.

We add together all three interactions. With the weights $r_{1}=1$,
$r_{2}=K_{2}/K_{1}$ and $r_{3}=K_{3}/K_{1}$, we get six equations:
\begin{ea}
\Omega^{2} Q_{A\rho}&=&[(\sum_{i=1}^{3} 3u_{i}^{2}
s_{iz}^{2}) Q_{A\rho}+...]+r_{2}[(8\sum_{i=1}^{3} v_{i}^{2}
\widetilde{s_{iz}}^{2}C_{i}^{2}) Q_{A\rho}+...]\nonumber\\
&+&r_{3}[(c^{2}+|\widetilde{c}|^{2}) Q_{A\rho}+...],\;\;\;
\mbox{\textit{etc.}}\nonumber
\end{ea}

From these equations, we can calculate the correct phonon dispersion
of SWNTs. To varify our method, we calculate a zigzag SWNTs (10,0).
As ref. \cite{M2}, we set $\Omega =\sqrt{3}$ to the optical phonon
at $1600$ cm$^{-1}$. For this tube, the two acoustic modes have
velocities $v_{L}=16.4$ km/s and $v_{TW}=8.7$ km/s. The optical
phonons are at frequencies $1600$ cm$^{-1}$, $1580$ cm$^{-1}$,
$877$ cm$^{-1}$ and $291$ cm$^{-1}$. 

In summary, we consider the phonons in chiral SWNTs. Lower symmetry
of chiral SWNTs make the computation more difficult. We have developed
a computational method to deal with this difficulty and obtain the
reasonable results. It can be used to calculate the phonon dispersion
of arbitrary chiral SWNTs. From the phonon dispersion, we may do further
research on the phonons involved phenomena.

We are also grateful to Prof. G. D. Mahan for his suggestion of doing
research on chiral nanotubes, and Dr. G. S. Jeon for helpful discussion.
We thank Prof. J. Kong for modifying the manuscript, and Prof. Georgii
G. Samsonidge for his comments. 

.

\newpage
\newpage
\newpage
%%%%%%%%%%%%%%%%%%%%%%%%%%%%%%%%%%%%%%%%%%%%%%%%%%%%%%%%%%%%
\begin{figure}[http]
\scalebox{0.8}{\includegraphics{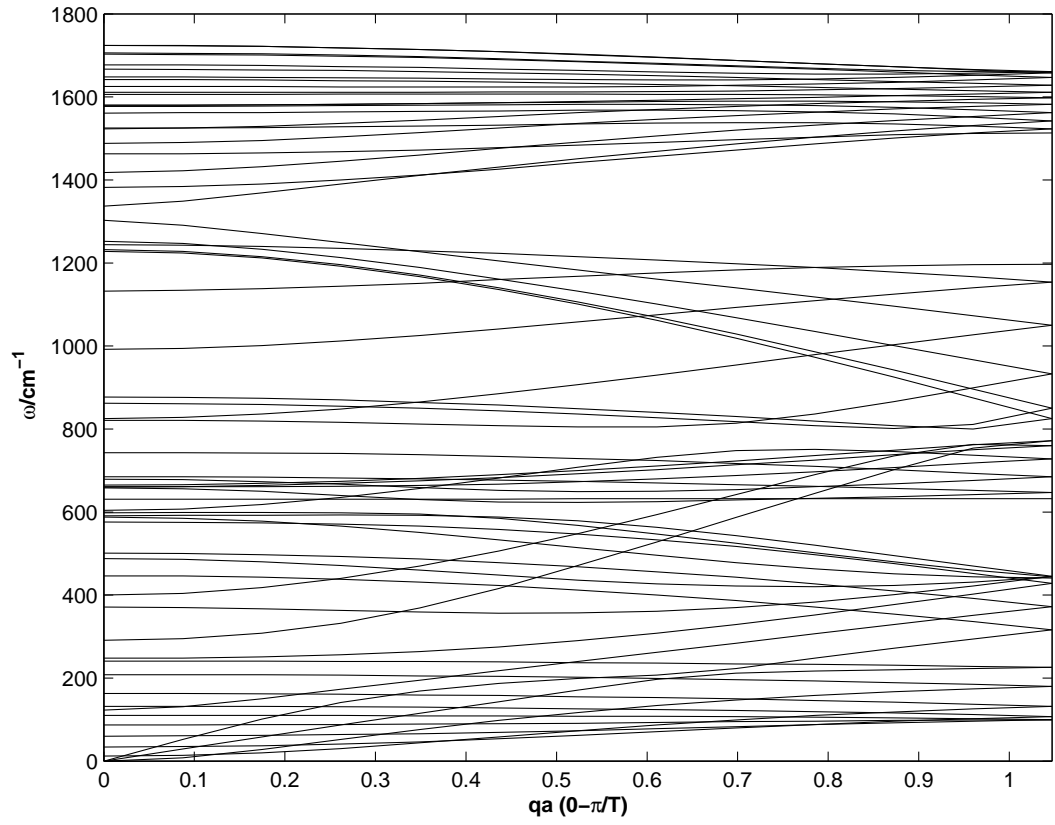}}
\caption{\label{Fig2} Phonons in a zigzag (10,0) tube.
$\alpha=0,1,2,...,N-1$.}
\end{figure}
%%%%%%%%%%%%%%%%%%%%%%%%%%%%%%%%%%%%%%%%%%%%%%%%%%%%%%%%%%%%
\newpage
\begin{figure}[http]
\scalebox{0.8}{\includegraphics{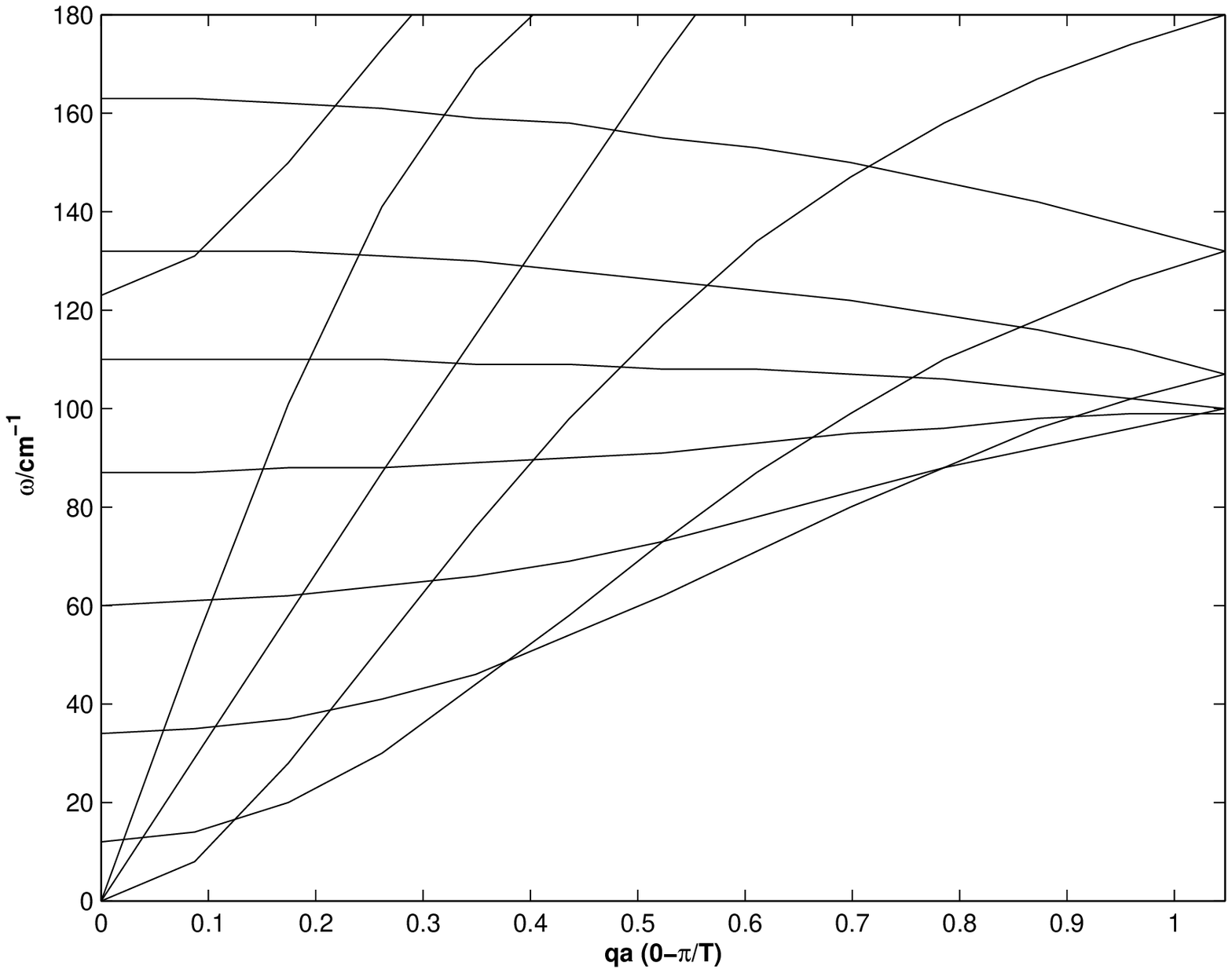}}
\caption{\label{Fig3} Large scale presentation of the region at
low frequency and small wave vector of (10,0) tube.}
\end{figure}
%%%%%%%%%%%%%%%%%%%%%%%%%%%%%%%%%%%%%%%%%%%%%%%%%%%%%%%%%%%

\begin{thebibliography}{10}
\bibitem{De}R. Saito, G. Dresselhaus and M. S. Dresselhaus, \textit{Physical Properties
of Carbon Nanotubes} (Imperial College Press, London, 1998). 
\bibitem{kong}B. J. LeRoy, S. G. Lemay, J. Kong and C. Dekker, Nature (London),
\textbf{432}, 371~(2004). 
\bibitem{Nick}A.N.Vamivakas, Y.Yin, A. G. Walsh, M.S. $\ddot{\mathrm{U}}$nl$\ddot{\mathrm{u}}$,
B. B. Goldberg, and A. K. Swan, submitting to Phys. Rev. B.
\bibitem{1}G. D. Mahan, Phys. Rev. B., \textbf{65}, 235402~(2002). 
\bibitem{Ye}L-H. Ye, B-G. Liu, D-S. Wang, and R. Han, Phys. Rev. B., \textbf{69},
235409~(2004).
\bibitem{popov}
Z. M. Li, V. N. Popov, and Z. K. Tang, Solid State Communications, \textbf{130}, 657~(2004).
\bibitem{Dobard}
E. Dobard$\check{z} \rm{i} \acute{c}$, I. Milo$\check{s}$evi$\acute{c}$, B. Nikoli$\acute{c}$, T. Vukovi$\acute{c}$, and M. Damnjanovi$\acute{c}$, Phys. Rev. B., \textbf{68},
045408~(2003).
\bibitem{Daniel}
D. S$\acute{a}$nchez-Portal, E. Artacho, J. M. Soler, A. Rubio, and P. Ordej$\acute{o}$n, Phys. Rev. B., \textbf{59}, 12678~(1999).
\bibitem{Mounet}
N. Mounet, and N. Marzari, Phys. Rev. B., \textbf{71}, 205214~(2005).
\bibitem{Dubay}
O. Dubay, and G. Kresse, Phys. Rev. B., \textbf{67}, 035401~(2003).
\bibitem{2}I. Milo$\breve{s}$evi$\acute{c}$, E. Dobard$\breve{z}$i$\acute{c}$
and M. Damnjanovic, Phys. Rev. B., \textbf{72}, 085426~(2005). 
\bibitem{3}C. Y. Wang, C. Q. Ru and A. Mioduchowski, Phys. Rev. B., \textbf{72},
075414~(2005). 
\bibitem{4}S. V. Goupalov, Phys. Rev. B., \textbf{71}, 085420~(2005). 
\bibitem{5}K. -P. Bohnen, R. Heid, H. J. Liu, and C. T. Chan, e-print: cond-mat/0411515. 
\bibitem{6}L. M. Woods and G. D. Mahan, Phys. Rev. B., \textbf{61}, 10651~(2000). 
\bibitem{M1}G. D. Mahan, Phys. Rev. B, \textbf{68}, 125409~(2003). 
\bibitem{M2}G. D. Mahan and G. S. Jeon, Phys. Rev. B, \textbf{70}, 075405~(2004). 
\bibitem{tu}Z. Tu, Z. C. Ou-Yang, J. Phys. C., 15, 6759 (2003); Phys. Rev.
B., \textbf{68}, 153403~(2003).
\bibitem{8}M. S. Dresselhaus and P. C. Eklund, Adv. Phys., \textbf{49}, 705~(2000). 
\bibitem{9}M. A. Stroscio and M. Dutta, \textit{Phonons in Nanostructures} (Cambridge
University Press, Cambridge, 2001). 
\end{thebibliography}
\end{document}